\documentclass[aps,prd,groupedaddress,preprintnumbers,
              floatfix, nofootinbib,longbibliography]{revtex4}
\usepackage{textcomp}
\usepackage[latin1]{inputenc}                    
\usepackage{graphicx}                            
\usepackage{epsfig}
\usepackage[export]{adjustbox}
\usepackage{mathrsfs}
\usepackage{mathtools}
\usepackage{slashed}

\usepackage{epsf}
\usepackage{latexsym}                            
\usepackage{amsfonts}                            
\usepackage{amssymb}                             
\usepackage{amsmath}                             
\usepackage{slashed}
\usepackage[mathscr]{eucal}                      
\usepackage{dcolumn}                             
\usepackage{multirow}                               
\usepackage{bm}                                  
\usepackage{hyperref}                            
\usepackage[usenames]{color}
\usepackage[dvipsnames]{xcolor}
\usepackage{ulem}

\graphicspath{{.}{Graphics/}}                    
\newcommand{\msbar}{\overline{\mbox{\rm MS}}}  
\newcommand{\mmsbar}{{\rm M}\overline{\mbox{\rm MS}}}  

\newcommand{\be}{\begin{equation}}
\newcommand{\ee}{\end{equation}}
\newcommand{\bea}{\begin{eqnarray}}
\newcommand{\eea}{\end{eqnarray}}

\def \3{\ss }

\newcommand{\beqn}{\begin{eqnarray}}
\newcommand{\eeqn}{\end{eqnarray}}

\newcommand{\UV}{\text{\tiny{UV}}}
\newcommand{\IR}{\text{\tiny{IR}}}

\def\poz{b}
\def\teml{a}
\def\bon{c}

\begin{document}

\begin{titlepage}
  {\vspace{-0.5cm} \normalsize
  \hfill \parbox{60mm}{
}}\\[10mm]
  \begin{center}
    \begin{LARGE}
            \textbf{One-loop matching for the twist-3 parton distribution $g_T (x)$}
    \end{LARGE}
  \end{center}

\vspace*{1cm}

 \vspace{-0.8cm}
  \baselineskip 20pt plus 2pt minus 2pt
  \begin{center}
    \textbf{
      Shohini Bhattacharya$^{(\teml)}$,
      Krzysztof Cichy$^{(\poz)}$,
      Martha Constantinou$^{(\teml)}$,\\
      Andreas Metz$^{(\teml)}$,
      Aurora Scapellato$^{(\poz)}$,
      Fernanda Steffens$^{(\bon)}$
      }
\end{center}

  \begin{center}
    \begin{footnotesize}
      \noindent 	
    $^{(\teml)}$ {\it Temple University, 1925 N.~12th Street, Philadelphia, PA 19122-1801, USA} \\
 	$^{(\poz)}$  {\it Faculty of Physics, Adam Mickiewicz University, Uniwersytetu Pozna\'{n}skiego 2, 61-614 Pozna\'{n}, Poland} \\
 	$^{(\bon)}$ {\it Institut f\"{u}r Strahlen- und Kernphysik, Rheinische Friedrich-Wilhelms-Universit\"{a}t Bonn,  \\ 
 	 Nussallee 14-16, 53115 Bonn, Germany} \\
     \vspace{0.2cm}
    \end{footnotesize}
  \end{center}

\centerline{\today}

\begin{abstract}
Perturbative matching relates the parton quasi-distributions, defined by Euclidean correlators at finite hadron momenta, to the light-cone distributions which are accessible in experiments. 
Previous matching calculations have exclusively focused on twist-2 distributions.
In this work, we address, for the first time, the one-loop matching for the twist-3 parton distribution function $g_T(x)$. 
The results have been obtained using three different infrared regulators, while dimensional regularization has been adopted to deal with the ultraviolet divergences. 
We present the renormalized expressions of the matching coefficient for $g_{T}(x)$ in the $\overline{\mbox{\rm MS}}$ and modified $\overline{\mbox{\rm MS}}$ schemes. 
We also discuss the role played by a zero-mode contribution. 
Our results have already been used for the extraction of $g_T(x)$ from lattice QCD calculations.
\end{abstract}

\maketitle
\end{titlepage}

\section{Introduction}
\label{s:introduction}
Parton distribution functions (PDFs) are important quantities encoding information about spatial and momentum distributions of partons inside hadrons~\cite{Collins:1981uw}. 
PDFs are non-perturbative objects which can be extracted from data on high-energy scattering experiments by making use of factorization theorems in quantum chromodynamics (QCD)~\cite{Collins:1989gx}.
Additional input on PDFs has been obtained through a variety of model calculations and, in particular, \textit{ab initio} calculations within the framework of lattice QCD~\cite{Lin:2017snn,Cichy:2018mum}. 
PDFs are light-cone dominated, which corresponds to a single point in a 4-dimensional Euclidean space-time in which lattice calculations are perfomed. 
Therefore, until recently, lattice QCD computations of PDFs were limited to matrix elements of local operators which are related to Mellin moments of PDFs. 
These only give partial information on PDFs, and practically, only the first two nontrivial moments can be studied reliably. Consequently, the full dependence of PDFs on the parton momentum fraction $x$ remained inaccessible.

In 2013, Ji proposed to address the $x$-dependence of PDFs in lattice QCD by means of so-called parton quasi-distributions (quasi-PDFs)~\cite{Ji:2013dva,Ji:2014gla}.
Quasi-PDFs are defined in terms of matrix elements of equal-time non-local operators which are purely spacelike and therefore can be readily computed on Euclidean lattices.
The matrix elements are considered for hadron states with momentum in a given direction, say $P=(P^0,0,0,P^3)$, and the quasi-PDFs reduce to their corresponding standard (light-cone) PDFs when $P^{3} = |\vec{P}| \rightarrow \infty$ prior to renormalization of ultraviolet (UV) divergences. 
In lattice computations, UV cut-offs ($\Lambda$) are given by the finite lattice spacing $a$ ($\Lambda \sim a^{-1}$), and one (naturally) deals with UV renormalization before taking the limit $P^{3} \rightarrow \infty$. 
The limits $\Lambda \rightarrow \infty$ and $P^{3} \rightarrow \infty$ do not commute, which leads to non-trivial differences in the UV behavior of the quasi-PDFs and light-cone PDFs.
On the other hand, the essence of the approach is the fact that the quasi-PDFs and light-cone PDFs share the same non-perturbative physics~\cite{Ji:2013dva,Ji:2014gla}, while the UV disparities can be systematically computed through a perturbative procedure known as matching~\cite{Xiong:2013bka}.
It was pointed out that a matching formula for quasi-PDFs is basically equivalent to the aforementioned factorization of hard scattering cross sections --- the quasi-PDF ``lattice observable" is factorized into a PDF and perturbatively calculable matching coefficients~\cite{Ma:2014jla}.
All-order factorization for (twist-2) quasi-PDFs has been addressed for the first time in Ref.~\cite{Ma:2014jla} by analyzing Feynman diagrams.
Other studies of factorization made use of the operator product expansion --- see, for instance, Ref.~\cite{Izubuchi:2018srq}.
We also mention that other approaches for addressing the $x$-dependence of PDFs have been suggested~\cite{Braun:1994jq, Detmold:2005gg, Braun:2007wv, Ma:2014jla, Chambers:2017dov, Hansen:2017mnd, Radyushkin:2017cyf, Orginos:2017kos, Ma:2017pxb, Radyushkin:2017lvu, Liang:2017mye, Detmold:2018kwu, Radyushkin:2019mye}, with some of them 
closely related to the quasi-PDFs.

By now, various aspects of quasi-PDFs and related quantities have been studied in detail~\cite{Radyushkin:2017ffo, Carlson:2017gpk, Briceno:2017cpo, Constantinou:2017sej, Alexandrou:2017huk, Chen:2017mzz, Rossi:2017muf, Ji:2017rah, Ji:2017oey, Ishikawa:2017faj, Green:2017xeu, Chen:2017mie, Monahan:2017hpu, Ji:2018hvs, Xu:2018mpf, Jia:2018qee, Briceno:2018lfj, Spanoudes:2018zya, Radyushkin:2018nbf, Ji:2018waw, Karpie:2018zaz, Zhang:2018diq, Li:2018tpe, Braun:2018brg, Ebert:2018gzl, Ebert:2019okf, Constantinou:2019vyb, Karpie:2019eiq, Zhang:2019qiq, Ebert:2019tvc, Ji:2019sxk, Shanahan:2019zcq, Ji:2019ewn, Huo:2019vdl, Vladimirov:2020ofp, Green:2020xco, Braun:2020ymy, Ebert:2020gxr}. 
In particular, a considerable number of pioneering lattice QCD results has been obtained in the meantime --- see Refs.~\cite{Alexandrou:2018pbm, Chen:2018xof, Chen:2018fwa, Alexandrou:2018eet, Liu:2018uuj, Bali:2018spj, Lin:2018pvv, Fan:2018dxu, Liu:2018hxv, Sufian:2019bol, Alexandrou:2019lfo, Izubuchi:2019lyk, Chen:2019lcm, Liang:2019frk, Cichy:2019ebf, Joo:2019jct, Joo:2019bzr, Sufian:2020vzb, Chai:2020nxw, Shanahan:2020zxr, Lin:2020ssv, Joo:2020spy, Bhattacharya:2020cen, Zhang:2020dkn, Bhat:2020ktg} for recent work. 
Moreover, the properties of quasi-PDFs have been explored in several models~\cite{Gamberg:2014zwa, Bacchetta:2016zjm, Nam:2017gzm, Broniowski:2017wbr, Hobbs:2017xtq, Broniowski:2017gfp, Xu:2018eii, Bhattacharya:2018zxi, Bhattacharya:2019cme, Son:2019ghf, Ma:2019agv, Kock:2020frx, Luo:2020yqj}. 
The tremendous progress in this field has recently been reviewed in Refs.~\cite{Cichy:2018mum, Ji:2020ect}.

The perturbative matching framework has already been explored extensively for twist-2 parton correlation functions~\cite{Xiong:2013bka, Ma:2014jla, Ji:2015jwa, Ji:2015qla, Xiong:2015nua, Wang:2017qyg, Stewart:2017tvs, Radyushkin:2018cvn, Zhang:2018ggy, Izubuchi:2018srq, Liu:2018tox, Liu:2019urm, Wang:2019tgg, Wang:2019msf, Radyushkin:2019owq, Balitsky:2019krf}. 
Despite such a commendable progress, so far no information is available about matching for higher-twist parton correlation functions, which include the three (two-parton) twist-3 PDFs $e$, $g_{T}$ and $h_{L}$ that exist for spin-$\frac{1}{2}$ hadrons such as the nucleon~\cite{Jaffe:1991kp, Jaffe:1991ra}.
Twist-3 PDFs do not have a density interpretation, in contrast to twist-2 PDFs.
However, they contain new information about quark-gluon-quark correlations inside a hadron~\cite{Balitsky:1987bk, Kanazawa:2015ajw}.
Moreover, twist-3 PDFs are not necessarily smaller than twist-2 PDFs.
On the other hand, it is difficult to measure twist-3 PDFs because they typically suffer from a kinematical suppression in physical observables.
Our focus here is on $g_T$, which, among the aforementioned three twist-3 PDFs, has received most of the attention so far.
It appears in the ``simple" inclusive deep-inelastic lepton-nucleon scattering process (DIS)~\cite{Jaffe:1996zw}, while $e$ and $h_L$ decouple from DIS.
We refer to~\cite{Flay:2016wie, Armstrong:2018xgk} for related recent measurements and to~\cite{Accardi:2009au, Sato:2016tuz} for attempts to extract information on $g_T$ from experimental data.

At the twist-2 level, one-loop matching relations between light-cone PDFs and quasi-PDFs are obtained by computing both quantities for a quark target in perturbative QCD (pQCD). 
In the present work, we aim at extending this very method to the twist-3 PDF $g_T$.
Partial computations of light-cone twist-3 PDFs for a quark target in pQCD can be found in the literature --- see for instance Refs.~\cite{Kundu:2001pk, Burkardt:2001iy, Aslan:2018tff}.
Here we extend the calculation for $g_T$ by going beyond the UV-divergent pieces and by computing the corresponding quasi-PDF.
We find that, diagram by diagram, the IR poles of the light-cone PDF $g_T$ and the quasi-PDF $g_{T,\rm{Q}}$ exactly match.
This key feature of the quasi-PDF approach is shown here explicitly for the first time for a twist-3 parton correlator, and encourages us to come up with a matching formula which we have used very recently for the first calculation of $g_T(x)$ in lattice QCD~\cite{Bhattacharya:2020cen}.

We regulate the UV divergences using dimensional regularization (DR).
In the non-perturbative (infrared (IR)) region, we exploit three different regulators: nonzero gluon mass, nonzero quark mass, and DR.
These regulators have been used before for the calculation of matching relations of twist-2 operators, and whenever results for different regulators were explicitly compared, it was reported that the final matching coefficient is regulator-independent.
We observe that, in our approach, for $g_T$ the situation is slightly more complicated due to a $\delta(x)$ zero-mode contribution.
It turns out that the IR divergence related to this term, strictly speaking, cannot be regulated using a nonzero gluon mass. 
Treating this term in isolation, and regulating it with either a nonzero quark mass or DR, we find that eventually (again) all three regulators provide the same matching coefficient.
Note that delta function zero-mode contributions have already received some attention in relation to twist-3 PDFs, while they are generally believed to be absent for twist-2 PDFs --- see~\cite{Burkardt:2001iy, Efremov:2002qh, Wakamatsu:2003uu, Aslan:2018zzk, Pasquini:2018oyz, Aslan:2018tff, Ma:2020kjz, Ji:2020baz} and references therein.

We organize the manuscript as follows: In Sec.~\ref{s:definitions}, we recall the definition of the light-cone PDF $g_{T}(x)$ and specify the corresponding quasi-PDF $g_{T, \rm Q} (x)$. Sec.~\ref{s:1LoopResults} is dedicated to presenting the one-loop pQCD results for $g_T (x)$ in the Feynman gauge using the three IR regulators. 
The (renormalized) matching kernel is derived in Sec.~\ref{s:MatchingKernel} in two schemes: the $\msbar$ scheme and the recently proposed modified $\msbar$ ($\mmsbar$) scheme~\cite{Alexandrou:2019lfo}. Qualitative differences in the matching formula depending upon these two schemes have been discussed as well. 
In Sec. \ref{s:summary}, we summarize our work and present a brief outlook.

\section{Definitions}
\label{s:definitions}
The light-cone PDF $g_T(x)$ of the nucleon is defined via (see for instance Ref.~\cite{Jaffe:1996zw})\footnote{For a generic four-vector $v$ we denote the Minkowski components by $(v^0, v^1, v^2, v^3)$ and the light-cone components by $(v^+, v^-, \vec{v}_\perp)$, with $v^+ = \frac{1}{\sqrt{2}} (v^0 + v^3)$, $v^- = \frac{1}{\sqrt{2}} (v^0 - v^3)$ and $\vec{v}_\perp = (v^1, v^2)$. Note also that in Eq.~(\ref{e:gT_def}) we suppress a flavor index.}
\begin{equation}
\frac{M}{P^+} \, S_\perp^i \, g_T(x) = 
\int \frac{dz^-}{4\pi} \, e^{i k \cdot z} \, \langle P, S | \bar{\psi}(- \tfrac{z}{2}) \, \gamma_\perp^i \gamma_5 \, {\cal W}(- \tfrac{z}{2}, \tfrac{z}{2}) \,\psi(\tfrac{z}{2})  | P, S \rangle \Big|_{z^+ = 0, \vec{z}_\perp = \vec{0}_\perp} \,,
\label{e:gT_def}
\end{equation}
where $M$ is the mass of the nucleon, $P$ its 4-momentum, `$i$' a transverse index, and $\gamma_5$ the Dirac matrix which anti-commutes with all the other gamma matrices.
The covariant spin vector $S$ of the nucleon is given by 
\begin{equation}
S^{\mu} = (S^+, S^-, \vec{S}_\perp) = \bigg( \lambda \frac{P^+}{M}, - \lambda \frac{M}{2 P^+}, \vec{S}_\perp \bigg) 
= \frac{1}{2M} \, \bar{u}(P,S) \, \gamma^\mu \gamma_5 \, u(P,S) \,,
\end{equation}
with $\lambda$ the nucleon helicity, and $u(P,S)$ the momentum space Dirac spinor of the nucleon.
Color gauge invariance of the bilocal quark operator in Eq.~(\ref{e:gT_def}) is ensured by the Wilson line
\begin{equation}
{\cal W}(- \tfrac{z}{2}, \tfrac{z}{2}) \Big|_{z^+ = 0, \vec{z}_\perp = \vec{0}_\perp}
= {\cal P} \, \textrm{exp} \, \Bigg( - i g_s \int_{-\tfrac{z^-}{2}}^{\tfrac{z^-}{2}} \, dy^- \, A^+(0^+, y^-, \vec{0}_\perp) \Bigg) \,,
\end{equation}
where ${\cal P}$ indicates path-ordering, $g_s$ the strong coupling constant, and $A^+$ the plus-component of the gluon field.

The quasi-PDF $g_{T,{\rm Q}}$, on the other hand, can be defined through a spatial correlation function according to~\cite{Ji:2013dva, Bhattacharya:2018zxi}
\begin{equation}
\label{e:gTQ_def}
\frac{M}{P^{3}} \, S_\perp^i \, g_{T, {\rm Q}}(x; P^{3})
= \int \frac{dz^3}{4\pi} \, e^{i k \cdot z} \, \langle P, S | \bar{\psi}(- \tfrac{z}{2}) \, \gamma_\perp^i \gamma_5 \, {\cal W}_{\rm Q}(- \tfrac{z}{2}, \tfrac{z}{2}) \,\psi(\tfrac{z}{2})  | P, S \rangle \Big|_{z^0 = 0, \vec{z}_\perp = \vec{0}_\perp} \,,
\end{equation}
with the Wilson line
\begin{equation}
{\cal W}_{\rm Q}(- \tfrac{z}{2}, \tfrac{z}{2}) \Big|_{z^0 = 0, \vec{z}_\perp = \vec{0}_\perp}
= {\cal P} \, \textrm{exp} \, \Bigg( - i g_s \int_{-\tfrac{z^3}{2}}^{\tfrac{z^3}{2}} \, dy^3 \, A^3(0, \vec{0}_\perp, y^3) \Bigg) \,.
\end{equation}
The momentum fraction of the quark in Eq.~(\ref{e:gTQ_def}) is given by $x = \frac{k^3}{P^3}$, which, for finite hadron momenta, differs from the momentum fraction $\frac{k^+}{P^+}$ used in the light-cone PDF in Eq.~(\ref{e:gT_def}).
In general, quasi-PDFs have an explicit dependence on the hadron momentum $P^3$.
However, the definition of $g_{T,\rm{Q}}$ in Eq.~(\ref{e:gTQ_def}) is such that the $P^3$-dependence drops out when taking the lowest Mellin moment (see also Ref.~\cite{Bhattacharya:2019cme}),
\begin{equation}
\int dx \, g_{T,\rm{Q}}(x;P^3) = \int dx \, g_T(x) \,.
\end{equation}
This feature can help to check the systematics of lattice calculations.

\section{One-loop results for $g_T$}
\label{s:1LoopResults}
\begin{figure}[t]
\begin{center}
    \includegraphics[width=15cm]{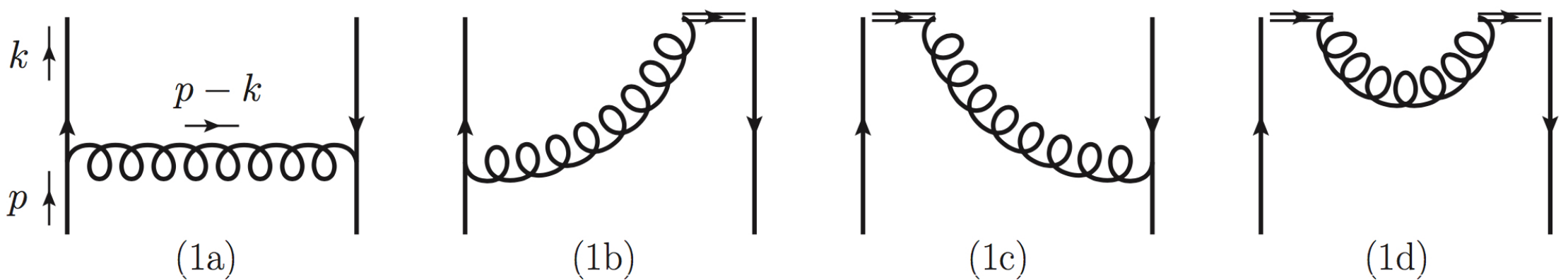}
 	\caption{One-loop real diagrams contributing to both $g_T$ and $g_{T,{\rm Q}}$.}
 	\label{fig:real}
\end{center}
\end{figure}

\begin{figure}[t]
\begin{center}
\hspace{0.5cm}
    \includegraphics[width=14.6cm]{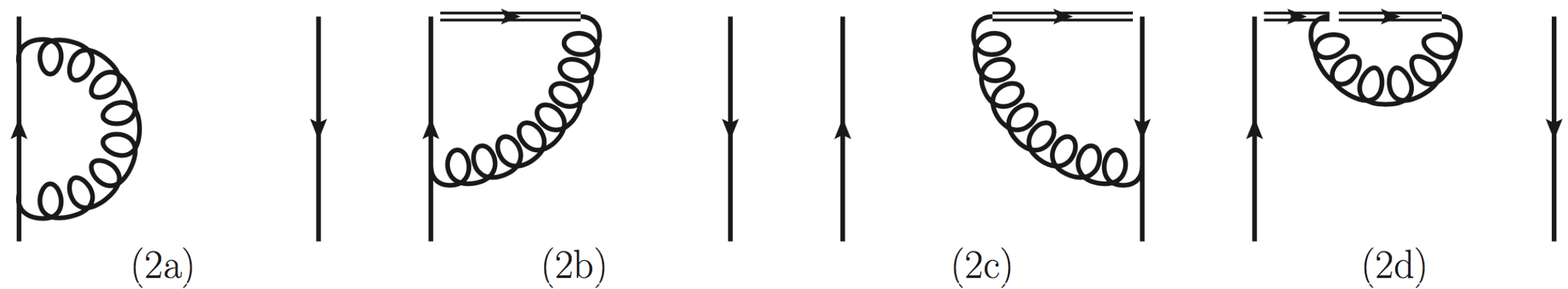}
 	\caption{One-loop virtual diagrams contributing to both $g_T$ and $g_{T,{\rm Q}}$.
 	The Hermitean conjugate diagrams of (2a) and (2d) are not shown.}
 	\label{fig:virtual}
\end{center}
\end{figure}
In this section, we compute the one-loop perturbative corrections for the light-cone PDF $g_T$ and the quasi-PDF $g_{T, \rm Q}$. 
We work in the Feynman gauge for which the one-loop real corrections are given by the diagrams (1a)--(1d) shown in Fig.~\ref{fig:real}, while the virtual corrections are given by diagrams (2a)--(2d) in Fig.~\ref{fig:virtual}. 
We will present our results using three different infrared (IR) regulators: nonzero gluon mass $m_g$, nonzero quark mass $m_q$, and dimensional regularization~(DR). 
For the ultraviolet (UV) divergences, DR will be used throughout the paper. 
Since we work at the parton level, the target mass will be denoted by $m_q$, the target spin vector by $s$, and its momentum by $p = x P$, where $x$ is the momentum fraction carried by the parton.

\subsection{Light-cone PDF}
First, we take up the real diagrams. For Fig.~(1a), the one-loop correction is
\begin{eqnarray}
\dfrac{m_{q}s^{i}_{\perp}}{p^{+}}\,g^{\rm{(1a)}}_{T} (x) = -\dfrac{i g^{2} C_{F} \mu^{2\epsilon} g_{\mu \nu}}{4} \int^{\infty}_{-\infty} \dfrac{d^{n}k}{(2\pi)^{n}} \dfrac{{\rm Tr} \big [ (\slashed{p}+ m_{q}) (1+\gamma_{5}\slashed{s}) \, \gamma^{\nu} \, (\slashed{k}+m_{q}) \, \gamma^{i}_{\perp}\gamma_{5} \, (\slashed{k}+m_{q}) \, \gamma^{\mu} \big ]}{(k^{2} -m^{2}_{q}+ i\varepsilon)^{2} ((p-k)^{2}-m^{2}_{g} + i\varepsilon)} \, \delta \bigg ( x- \dfrac{k^{+}}{p^{+}} \bigg ) \dfrac{1}{p^{+}} \,, \;\; 
\label{e:diagram_1a_LC}
\end{eqnarray}
where $g$ is the coupling constant associated with the quark-gluon vertex, $C_F = 4/3$ is the color factor and $n=4-2\epsilon$. 
In general, we use $\epsilon$ as the DR regulator, noticing that $\epsilon \rightarrow \epsilon_\UV > 0$ (with the corresponding subtraction scale as $\mu \rightarrow \mu_\UV > 0$) if it is used for the UV region, while $\epsilon \rightarrow \epsilon_\IR < 0$ if it is used for the IR region (with $\mu \rightarrow \mu_\IR > 0$).
From the definition of $g_T$ it is clear that the quark mass cannot be set to zero from the start, but must be kept finite until we have extracted the terms which are linear in $m_q$. 
After this step, one can set $m_q=0$ unless it is used as the IR regulator.
With this in mind, Eq.~(\ref{e:diagram_1a_LC}) can be simplified to
\begin{equation}
g^{\rm{(1a)}}_{T} (x) = -\dfrac{i g^{2} C_{F} \mu^{2\epsilon}}{(2\pi)^n} p^+ \int^{\infty}_{-\infty} d^{n-2}k_\perp d k^-  dk^+
\dfrac{(4-n) 2 \, p\cdot k -(2-n) (k^{2}+m^{2}_{q}) + 2 k^2_\perp} 
{(k^{2} -m^{2}_{q}+ i\varepsilon)^{2} ((p-k)^{2}-m^{2}_{g} + i\varepsilon)} \, \delta  \bigg ( x- \dfrac{k^{+}}{p^{+}} \bigg ) \dfrac{1}{p^{+}} \,.
\label{e:gT_1a_LC}
\end{equation}
In the following, we use the abbreviation 
\begin{displaymath}
{\cal P}_{\UV} = \frac{1}{\epsilon_{\UV}} + \ln 4\pi - \gamma_E \,,
\end{displaymath}
and likewise for the IR.
After performing the momentum integrals in Eq.~(\ref{e:gT_1a_LC}), one gets
\begin{equation}
g_T^{\rm{(1a)}}(x) \Big |_{m_{g}} =  \frac{\alpha_s C_F}{2\pi}  \, \bigg( - \delta(x) + x \, {\cal P}_{\UV} + x \ln \frac{\mu_{\UV}^2}{x m_g^2} + (1 - x) \bigg) \, ,
\label{e:gT_1a_mg_LC}
\end{equation}
or
\begin{equation}
g_T^{\rm{(1a)}}(x) \Big |_{m_{q}} = \frac{\alpha_s C_F}{2\pi}  \, \bigg( - \delta(x) + x \, {\cal P}_{\UV} +
x \ln \frac{\mu_{\UV}^2}{(1-x)^2 m_q^2} + \frac{x^2-2x-1}{1 - x} \bigg) \, ,
\end{equation}
depending on whether the gluon or quark mass is used as the IR regulator. 
Both results, however, have contributions from the point $x=0$ reflected through the Dirac delta function. 
To see the origin of this term, we notice that one of the momentum integrals in Eq.~(\ref{e:gT_1a_LC}) can be evaluated as (see also Ref.~\cite{Yan:1973qg})
\begin{equation}
(n-4)\int d k^- \frac{1}{(k^2 - m_q^2 + i\varepsilon)^{2}} = (n-4) \, \frac{i \pi}{k^2_\perp + m_q^2} \, \frac{\delta (x)}{p^{+}} \,,
\label{e:Dirac_delta}
\end{equation}
and, hence, the DR regulated UV pole (from the $k_{\perp}$ integral) makes a term proportional to $\delta(x)$ contribute to $g_T$.
The analysis of Eq.~(\ref{e:Dirac_delta}) in the case of a nonzero gluon mass deserves extra discussion.
Strictly speaking, in this case one should set the quark mass to zero in Eq.~(\ref{e:Dirac_delta}). 
But then one is left with an IR singularity.
Therefore, $m_g \neq 0$ is not sufficient to regulate all IR divergences for the $g_T$ calculation.
Such a feature shows up for the first time at twist-3.
Terms like in Eq.~(\ref{e:Dirac_delta}) are not present in the twist-2 case and are related to the zero-mode contributions~\cite{Burkardt:2001iy, Efremov:2002qh, Wakamatsu:2003uu, Aslan:2018zzk, Pasquini:2018oyz, Aslan:2018tff, Ma:2020kjz, Ji:2020baz}. 
In the following we do not abandon the gluon mass regulator, but rather consider two pragmatic options: (a) retain the quark mass in Eq.~(\ref{e:Dirac_delta}), and (b) use DR, while for all other contributions we keep working with a finite $m_g$. 
The two routes lead to two different answers. 
Option~(a) gives rise to a $\delta(x)$ term in $g_T$, as explained above.
For Option~(b), the corresponding IR pole $1/\epsilon_{\IR}$ also allows for a $\delta(x)$ through Eq.~(\ref{e:Dirac_delta}), but with an opposite sign to the contribution from the UV pole. 
Hence, for DR for the IR region the delta-function singularity drops out in $g_T$. 
The expression in Eq.~(\ref{e:gT_1a_mg_LC}) corresponds to Option (a). 
We emphasize that this choice is not necessary, and we discuss the impact of the two options on the matching coefficient later on.

The result for diagram~(1a) using DR is
\begin{equation}
g_T^{\rm{(1a)}}(x) \Big |_{\epsilon_{\IR}} = \frac{\alpha_s C_F}{2\pi} \, \bigg(x \,  ({\cal P}_{\UV} - {\cal P}_{\IR} )+
x \ln \frac{\mu_{\UV}^2}{\mu_{\IR}^2} \bigg) \,,
\end{equation}
where $\mu_{\IR}$ is the subtraction scale associated to the IR pole.

For the diagram in Fig.~(1b), the momentum integral is
\begin{eqnarray}
\dfrac{m_{q}s^{i}_{\perp}}{p^{+}}\,g^{\rm{(1b)}}_{T} (x) = -\dfrac{i g^{2} C_{F} \mu^{2\epsilon} g_{\mu \nu} v^{\nu}}{4} \int^{\infty}_{-\infty} \dfrac{d^{n}k}{(2\pi)^{n}} \dfrac{{\rm Tr} \big [ (\slashed{p}+ m_{q}) (1+\gamma_{5}\slashed{s}) \, \gamma^{i}_{\perp}\gamma_{5} \, (\slashed{k}+m_{q}) \, \gamma^{\mu} \big ]}{(v \cdot (p-k)+i\varepsilon)(k^{2} -m^{2}_{q}+ i\varepsilon) ((p-k)^{2}-m^{2}_{g} + i\varepsilon)} \, \delta \bigg ( x- \dfrac{k^{+}}{p^{+}} \bigg ) \dfrac{1}{p^{+}} \,, \;\,
\end{eqnarray}
where $v$ is the light-cone vector $v^{\mu} = (0^{+}, 1^{-}, \vec{0}_{\perp})$ such that $v^{2} =0$ and $v \cdot a = a^{+}$ for any generic four-vector $a^{\mu}$. 
The results are
\begin{eqnarray}
g_T^{\rm{(1b)}}(x) \Big |_{m_{g}} & = & \frac{\alpha_s C_F}{2\pi} \, \frac{1 + x}{2(1-x)} \, \bigg ( {\cal P}_{\UV} + \ln \frac{\mu_{\UV}^2}{x m_g^2} \bigg ) \,, 
\label{e:gT_1b_mg_LC}
\\[0.2cm]
g_T^{\rm{(1b)}}(x) \Big |_{m_{q}} & = & \frac{\alpha_s C_F}{2\pi} \, \frac{1 + x}{2(1-x)} \, \bigg ( {\cal P}_{\UV} + \ln \frac{\mu_{\UV}^2}{(1-x)^2 m_q^2} \bigg ) \,,
\\[0.2cm]
g_T^{\rm{(1b)}}(x) \Big |_{\epsilon_{\IR}} & = & \frac{\alpha_s C_F}{2\pi} \, \frac{1 + x}{2(1-x)} \,  \bigg ( {\cal P}_{\UV} - {\cal P}_{\IR} + \ln \frac{\mu_{\UV}^2}{\mu_{\IR}^2} \bigg ) \,.
\end{eqnarray}
Diagram~(1c) gives the same result as~(1b), while diagram~(1d) gives no contribution since the result is proportional to $v^{2}$.

We now proceed to the computation of the virtual diagrams. 
Fig.~(2a) shows the contribution from the quark self-energy diagram which is independent of the particular PDF under consideration. 
The self-energy is given by
\begin{equation}
-i\Sigma (p) = C_{F} \int^{\infty}_{-\infty} \dfrac{d^{n}k}{(2\pi)^{n}} (-ig\mu^{\epsilon}\gamma^{\mu}) \dfrac{i (\slashed{k}+m_q)}{k^{2}-m_q^2+i\epsilon} (-ig\mu^{\epsilon}\gamma^{\nu}) \dfrac{-ig_{\mu \nu}}{(p-k)^{2}-m^{2}_{g}+i\epsilon} \,,
\end{equation}
and the contribution to $g_T$ from diagram~(2a) (plus its Hermitean conjugate) is
\begin{eqnarray}
g^{\rm{(2a)}}_T \Big |_{m_{g}} &=& \dfrac{\partial \Sigma (p)}{\partial \slashed{p}} \Big |_{m_{g}} =
-\dfrac{\alpha_{s} C_{F}}{2 \pi} \int_{0}^{1} dy \, y  \bigg ( {\cal P}_{\UV} + \ln \dfrac{\mu_{\UV}^{2}}{y m^{2}_{g} } -1 \bigg ) \,,
\label{e:self_energy_mg_LC}
\\[0.2cm]
g^{\rm{(2a)}}_T \Big |_{m_{q}} &=& \dfrac{\partial \Sigma (p)}{\partial \slashed{p}} \Big |_{m_{q}} =
-\dfrac{\alpha_{s} C_{F}}{2 \pi} \int_{0}^{1} dy \, (1-y)  \bigg ( {\cal P}_{\UV} + \ln \dfrac{\mu_{\UV}^{2}}{(1-y)^2 m_q^{2}} - \dfrac{1+y^2}{(1-y)^{2}} \bigg ) \,,
\\[0.2cm]
g^{\rm{(2a)}}_T \Big |_{\epsilon_{\IR}} &=& \dfrac{\partial \Sigma (p)}{\partial \slashed{p}} \Big |_{\epsilon_{\IR}}  =
-\dfrac{\alpha_{s} C_{F}}{2 \pi} \int_{0}^{1} dy \, y  \bigg (  {\cal P}_{\UV}- {\cal P}_{\IR}  + \ln \dfrac{\mu_{\UV}^{2}}{\mu_{\IR}^{2} } \bigg )
\end{eqnarray}
for the three IR regulators.

The diagram in Fig.~(2b) provides
\begin{eqnarray}
\dfrac{m_{q}s^{i}_{\perp}}{p^{+}}\,g^{\rm{(2b)}}_{T} &=& \dfrac{i g^{2} C_{F} \mu^{2\epsilon} g_{\mu \nu} v^{\nu}}{4} \int^{\infty}_{-\infty} \dfrac{d^{n}k}{(2\pi)^{n}} \dfrac{{\rm Tr} \big [ (\slashed{p}+ m_{q}) (1+\gamma_{5}\slashed{s}) \, \gamma^{i}_{\perp}\gamma_{5} \, (\slashed{k}+m_{q}) \, \gamma^{\mu} \big ]}{(v \cdot (p-k)+i\varepsilon)(k^{2} -m^{2}_{q}+ i\varepsilon) ((p-k)^{2}-m^{2}_{g} + i\varepsilon)} \,, 
\end{eqnarray}
which is exactly the same expression as its counterpart, Fig.~(1b), except for an overall minus sign. 
The overall sign is due to the reversed momentum flow in the Wilson line relative to Fig.~(1b). 
Likewise, Fig.~(2c) gives the same contribution as Fig.~(1c), while there is no contribution from Fig.~(2d). 
The results for the virtual diagrams have an overall prefactor of $\delta(1-x)$ which we have left out for simplicity of notation.

For the matching presented in the following sections, we will use the $\msbar$ renormalized expressions of these results.
Here we have not studied potential mixing with other operators under renormalization, but rather leave this topic for future work. 

\subsection{Quasi-PDF}

Quasi-PDFs are given by the Fourier transform of purely spatial matrix elements between hadronic states of finite momenta.
With this in mind, the quasi-PDF resulting from Fig.~(1a) is written as in Eq.~(\ref{e:diagram_1a_LC}) with the replacement $p^+ \rightarrow p^{3}$. 
After integrating over $k^0$ (using the residue theorem) and over $k_{\perp}$ we find\footnote{For convenience of notation, in our results we use that $p^{2}_{3} = (p^{3})^{2}$.}
\begin{eqnarray}
g_{T,\rm{Q}}^{\rm{(1a)}}(x) \Big |_{m_{g}} & = & \frac{\alpha_s C_F}{2\pi}
\begin{dcases}
x \, \ln \frac{x}{x - 1} - 1
& \quad x > 1  \\[0.2cm]
x \, \ln \frac{4(1 - x) p^{2}_{3}}{m_g^2} + 1 - 2x
& \quad 0 < x < 1 \\[0.2cm]
x \, \ln \frac{x - 1}{x} + 1
& \quad x < 0 \,,
\end{dcases}
\label{e:gT_1a_mg_quasi}
\end{eqnarray}
\begin{eqnarray}
g_{T,\rm{Q}}^{\rm{(1a)}}(x) \Big |_{m_{q}} & = & \frac{\alpha_s C_F}{2\pi}
\begin{dcases}
x \, \ln \frac{x}{x - 1} - 1
& \quad x > 1  \\[0.2cm]
x \, \ln \frac{4 x p^{2}_{3}}{(1-x) m_q^2} + 1 - 2x + \frac{2}{x-1}
& \quad 0 < x < 1 \\[0.2cm]
x \, \ln \frac{x - 1}{x} + 1
& \quad x < 0 \,,
\end{dcases}
\label{e:gT_1a_mq_quasi}
\end{eqnarray}
\begin{eqnarray}
g_{T,\rm{Q}}^{\rm{(1a)}}(x) \Big |_{\epsilon_{\IR}} 
&=& \frac{\alpha_s C_F}{2\pi}
\begin{dcases}
x \, \ln \frac{x}{x - 1} - 1
& \quad x > 1  \\[0.2cm]
\delta(x) + x \, \ln \frac{4 x (1-x) p^{2}_{3}}{\mu_{\IR}^2} -x - x \, {\cal P}_{\IR}
& \quad 0 < x < 1 \\[0.2cm]
x \, \ln \frac{x - 1}{x} + 1
& \quad x < 0 \, .
\end{dcases}
\label{e:gT_1a_DR_quasi}
\end{eqnarray} 
More details about the derivation of the singular term (delta function contribution) in Eq.~\eqref{e:gT_1a_DR_quasi} can be found in Refs.~\cite{Bhattacharya:2020jfj, Bhattacharya:2021grn}.

In contrast to light-cone PDFs, for quasi-PDFs the $k_{\perp}$ integrals for real-emission diagrams are UV finite.
However, UV poles appear when the results are integrated over the momentum fraction $x$. 
As is by now well known, the quasi-PDFs have support outside the region $0 \le x \le 1$. 
But IR poles appear in that ``physical'' region only.
Moreover, for all three regulators the poles agree with the ones we obtained for the corresponding contribution to the light-cone PDF.
In fact, we find such an exact match of IR poles, which is at the heart of the quasi-PDF approach, for all the one-loop diagrams.   

For Fig.~(1b), the corresponding results are
\begin{eqnarray}
g_{T,\rm{Q}}^{\rm{(1b)}}(x) \Big |_{m_{g}} & = & \frac{\alpha_s C_F}{2\pi} \, \frac{1 + x}{2 (1 - x)}
\begin{dcases}
\ln \frac{x}{x - 1}
& \quad x > 1  \\[0.2cm]
\ln \frac{4(1 - x) p^{2}_{3}}{m_g^2}
& \quad 0 < x < 1 \\[0.2cm]
\ln \frac{x - 1}{x}
& \quad x < 0 \,,
\end{dcases}
\label{e:gT_1b_mg_quasi}
\\[0.2cm]
g_{T,\rm{Q}}^{\rm{(1b)}}(x) \Big |_{m_{q}} & = & \frac{\alpha_s C_F}{2\pi} \, \frac{1 + x}{2 (1 - x)}
\begin{dcases}
\ln \frac{x}{x - 1}
& \quad x > 1  \\[0.2cm]
\ln \frac{4x p^{2}_{3}}{(1-x) m_q^2}
& \quad 0 < x < 1 \\[0.2cm]
\ln \frac{x - 1}{x}
& \quad x < 0 \,,
\end{dcases}
\\[0.2cm]
g_{T,\rm{Q}}^{\rm{(1b)}}(x) \Big |_{\epsilon_{\IR}} & = & \frac{\alpha_s C_F}{2\pi} \, \frac{1 + x}{2 (1 - x)}
\begin{dcases}
\ln \frac{x}{x - 1}
& \quad x > 1  \\[0.2cm]
\ln \frac{4x(1-x) p^{2}_{3}}{\mu_{\IR}^2}-{\cal P}_{\IR}
& \quad 0 < x < 1 \\[0.2cm]
\ln \frac{x - 1}{x}
& \quad x < 0 \,.
\end{dcases}
\end{eqnarray}
The contribution from Fig.~(1c) is the same as from Fig.~(1b).

Finally, the contribution from Fig.~(1d) is calculated as
\begin{eqnarray}
\dfrac{m_{q}s^{i}_{\perp}}{p^{3}}\,g^{\rm{(1d)}}_{T, \rm Q} (x) = -\dfrac{i g^{2} C_{F} \mu^{2\epsilon} g_{\mu \nu} v^{\mu} v^{\nu}}{4} \int^{\infty}_{-\infty} \dfrac{d^{n}k}{(2\pi)^{n}} \dfrac{{\rm Tr} \big [ (\slashed{p}+ m_{q}) (1+\gamma_{5}\slashed{s}) \, \gamma^{i}_{\perp}\gamma_{5} \big ]}{(v \cdot (p-k)+i\varepsilon)^{2} ((p-k)^{2}-m^{2}_{g} + i\varepsilon)} \, \delta \bigg ( x- \dfrac{k^{3}}{p^{3}} \bigg ) \dfrac{1}{p^{3}} \,, \;\; 
\end{eqnarray}
which results in
\begin{eqnarray}
g_{T,\rm{Q}}^{\rm{(1d)}}(x) & = & \frac{\alpha_s C_F}{2\pi}
\begin{dcases}
\frac{1}{1 - x}
& \quad x > 1  \\[0.2cm]
\frac{1}{x - 1}
& \quad 0 < x < 1 \\[0.2cm]
\frac{1}{x - 1}
& \quad x < 0
\end{dcases}
\label{e:gT_1d_quasi}
\end{eqnarray}
for all three IR regulators.
Note that this diagram provides the same result for other parton distributions such as the twist-2 unpolarized PDF.

We now compute the quasi-PDFs for the virtual diagrams in Figs.~(2a)--(2d). 
When integrated over all parton momenta, the self-energy diagram in Fig.~(2a) gives the same contribution for both light-cone PDFs and quasi-PDFs. 
Nevertheless, to make contact with the techniques used in the computation of the real diagrams for quasi-PDFs, we first integrate over the $k^0$ and $k_\perp$ components, leaving the integral over $k^3$ to be made at the end. 
In this case, using the gluon mass as the IR regulator, and DR for the UV divergence, which now appears in the integration over $y$ when using $k^3 = y p^3$, the result is
\begin{equation}
\begin{split}
g^{\rm{(2a)}}_{T, \rm Q} \Big |_{m_{g}} = \dfrac{\partial \Sigma (p)}{\partial \slashed{p}} \Big |_{m_{g}} = - \dfrac{\alpha_{s} C_{F}}{2\pi}(1-\epsilon_{\UV}) C(\epsilon_{\UV})
\bigg (\frac{p^3}{\mu_{\UV}} \bigg )^{-2\epsilon_{\UV}} \int dy
\begin{dcases}
y^{\boldsymbol{-} 2\epsilon_{\UV}} \bigg ( y \ln \dfrac{y}{y-1} - 1\bigg )
&  y > 1  \\[0.2cm]
y^{\boldsymbol{-} 2\epsilon_{\UV}} \bigg ( y \ln \dfrac{4(1-y)p^{2}_{3}}{m_g^2}  + 1 -2y \bigg )
&  0 < y < 1 \quad \\[0.2cm]
(-y)^{\boldsymbol{-} 2\epsilon_{\UV}} \bigg ( y \ln \dfrac{y-1}{y} + 1 \bigg )
&  y < 0 \,,
\end{dcases}
\label{e:mg_selfenergy_qPDF}
\end{split}
\end{equation}
where
\begin{equation}
C(\epsilon_{\UV})=\dfrac{\pi^{1/2-\epsilon_{\UV}}}{(2\pi)^{-2\epsilon_{\UV}}\Gamma[1/2-\epsilon_{\UV}]} \,.
\end{equation}
Notice that Eq.~(\ref{e:mg_selfenergy_qPDF}) differs from Eq.~(26) of Ref.~\cite{Wang:2017qyg}.
Upon integrating Eq.~(\ref{e:mg_selfenergy_qPDF}) we obtain
\begin{equation}
g^{\rm{(2a)}}_{T, \rm Q} \Big |_{m_{g}} = \dfrac{\partial \Sigma (p)}{\partial \slashed{p}} \Big |_{m_{g}} = 
-\dfrac{\alpha_{s} C_{F}}{2 \pi} \bigg ( -\dfrac{1}{4} +  \dfrac{1}{2} \ln \dfrac{\mu_{\UV}^{2}}{m^{2}_{g} } + \dfrac{1}{2} \, {\cal P}_{\UV} \bigg ) \,,
\label{e:mg_selfenergy}
\end{equation}
which agrees with  Eq.~(\ref{e:self_energy_mg_LC}) after the integration has been performed. 
This serves as important consistency check of our results for the self-energy graph. For the other two IR regulators we obtain
\begin{eqnarray}
g^{\rm{(2a)}}_{T, \rm Q} \Big |_{m_{q}} = \dfrac{\partial \Sigma (p)}{\partial \slashed{p}} \Big |_{m_{q}} & = & -\dfrac{\alpha_{s} C_{F}}{2\pi} C(\epsilon_{\UV})
\bigg (\frac{p^3}{\mu_{\UV}} \bigg )^{-2\epsilon_{\UV}} \int dy
\begin{dcases}
(1-\epsilon_{\UV}) \, y^{\boldsymbol{-} 2\epsilon_{\UV}} \bigg ((1-y) \ln \dfrac{y}{y-1} + 1\bigg )
& \, y > 1  \\[0.2cm]
y^{\boldsymbol{-} 2\epsilon_{\UV}} \bigg ( (1-\epsilon_{\UV}) (1 - y) \ln \dfrac{4 y p^{2}_{3}}{(1-y) m_q^2}  \\[0.2cm] 
- (1-\epsilon_{\UV}) \frac{2 y^2 - 5 y + 1}{1-y}  \\[0.2cm]
-\bigg (1-\dfrac{\epsilon_{\UV}}{2} \bigg ) \frac{4y}{1-y} \bigg )
& \, 0 < y < 1 \\[0.2cm]
(1-\epsilon_{\UV}) \, (-y)^{\boldsymbol{-} 2\epsilon_{\UV}} \bigg ((1-y) \ln \dfrac{y-1}{y} - 1 \bigg )
& \, y < 0 \,,
\end{dcases}
\label{e:mq_self_energy_qPDF}
\end{eqnarray}
\begin{eqnarray}
g^{\rm{(2a)}}_{T, \rm Q} \Big |_{\epsilon_{\IR}} = \dfrac{\partial \Sigma (p)}{\partial \slashed{p}} \Big |_{\epsilon_{\IR}} & = & -\dfrac{\alpha_{s} C_{F}}{2\pi}(1-\epsilon_{\UV}) C(\epsilon_{\UV})
\bigg (\frac{p^3}{\mu_{\UV}} \bigg )^{-2\epsilon_{\UV}} \int dy
\begin{dcases}
y^{\boldsymbol{-} 2\epsilon_{\UV}} \bigg ( y \ln \dfrac{y}{y-1} - 1\bigg )
& \, y > 1  \\[0.2cm]
y^{\boldsymbol{-} 2\epsilon_{\UV}} \bigg ( y \ln \dfrac{4y(1-y) p^{2}_{3}}{\mu_{\IR}^2} \\[0.2cm] 
+ \, 1 - y - y \, {\cal P}_{\IR} \bigg )
& \, 0 < y < 1  \\[0.2cm]
(-y)^{\boldsymbol{-} 2\epsilon_{\UV}} \bigg ( y \ln \dfrac{y-1}{y} + 1 \bigg )
& \, y < 0 \,.
\end{dcases}
\label{e:self_energy_qPDF}
\end{eqnarray}
Once again, it is straightforward to show that Eqs.~(\ref{e:mq_self_energy_qPDF}) and~(\ref{e:self_energy_qPDF}) consistently reproduce their corresponding light-cone results with the appropriate IR regulators.

The results for the digrams in Fig.~(2b) (and Fig.~(2c)) read
\begin{eqnarray}
g^{\rm{(2b)}}_{T, \rm Q} \Big |_{m_{g}} & = & -\dfrac{\alpha_{s} C_{F}}{2\pi}
C(\epsilon_{\UV}) \bigg (\frac{p^3}{\mu_{\UV}} \bigg )^{-2\epsilon_{\UV}}
\int dy \frac{1+y}{2(1-y)}
\begin{dcases}
y^{\boldsymbol{-} 2\epsilon_{\UV}} \ln \frac{y}{y - 1}
& \quad y > 1  \\[0.2cm]
y^{\boldsymbol{-} 2\epsilon_{\UV}} \ln \frac{4(1 - y) p^{2}_{3}}{m_g^2}
& \quad 0 < y < 1 \\[0.2cm]
(-y)^{\boldsymbol{-} 2\epsilon_{\UV}} \ln \frac{y - 1}{y}
& \quad y < 0 \,,
\end{dcases}
\label{e:gT_2b_mg_quasi}
\\[0.2cm]
g^{\rm{(2b)}}_{T, \rm Q} \Big |_{m_{q}} & = & -\dfrac{\alpha_{s} C_{F}}{2\pi}
C(\epsilon_{\UV}) \bigg (\frac{p^3}{\mu_{\UV}} \bigg )^{-2\epsilon_{\UV}}
\int dy \frac{1+y}{2(1-y)}
\begin{dcases}
y^{\boldsymbol{-} 2\epsilon_{\UV}} \ln \frac{y}{y - 1}
& \quad y > 1  \\[0.2cm]
y^{\boldsymbol{-} 2\epsilon_{\UV}} \ln \frac{4 y p^{2}_{3}}{(1-y)m_q^2}
& \quad 0 < y < 1 \\[0.2cm]
(-y)^{\boldsymbol{-} 2\epsilon_{\UV}} \ln \frac{y - 1}{y}
& \quad y < 0 \,,
\end{dcases}
\label{e:gT_2b_mq_quasi}
\\[0.2cm]
\label{e:gT_2b_DR}
g^{\rm{(2b)}}_{T, \rm Q} \Big |_{\epsilon_{\IR}} & = &  -\dfrac{\alpha_{s} C_{F}}{2\pi}
C(\epsilon_{\UV}) \bigg (\frac{p^3}{\mu_{\UV}} \bigg )^{-2\epsilon_{\UV}}
\int dy \frac{1+y}{2(1-y)}
\begin{dcases}
y^{\boldsymbol{-} 2\epsilon_{\UV}} \ln \frac{y}{y - 1}
&  y > 1  \\[0.2cm]
y^{\boldsymbol{-} 2\epsilon_{\UV}} \bigg (\ln \frac{4y(1 - y) p^{2}_{3}}{\mu_{\IR}^2} - {\cal P}_{\IR} \bigg )
&  0 < y < 1 \\[0.2cm]
(-y)^{\boldsymbol{-} 2\epsilon_{\UV}} \ln \frac{y - 1}{y}
&  y < 0 \,,
\end{dcases}
\end{eqnarray}
for the three different IR regulators.

Finally, the result for the diagram in Fig.~(2d) is given by
\begin{eqnarray}
\label{gT_2d_quasi}
g^{\rm{(2d)}}_{T, \rm Q} =  -\dfrac{\alpha_{s} C_{F}}{2\pi}
C(\epsilon_{\UV}) \bigg (\frac{p^3}{\mu_{\UV}} \bigg )^{-2\epsilon_{\UV}}
\int dy 
\begin{dcases}
y^{\boldsymbol{-} 2\epsilon_{\UV}}  \frac{1}{1-y}
& \quad y > 1  \\[0.2cm]
y^{\boldsymbol{-} 2\epsilon_{\UV}} \frac{1}{y-1}
& \quad 0 < y < 1 \\[0.2cm]
(-y)^{\boldsymbol{-} 2\epsilon_{\UV}} \frac{1}{y-1}
& \quad y < 0 \quad .
\end{dcases} 
\end{eqnarray}
Strictly speaking, what we call here $g^{\rm{(2d)}}_{T, \rm Q}$ includes also its Hermitean conjugate.

\section{Matching Kernel}
\label{s:MatchingKernel}
%
Quasi-PDFs can be related to light-cone PDFs through a perturbatively calculable matching coefficient up to power corrections that are suppressed in the hadron momentum. 
Omitting the scale dependence, a corresponding matching formula schematically reads
\begin{equation}
\tilde{q} (x; P^{3}) = \int^{+1}_{-1} \dfrac{d y}{|y|} C \bigg ( \dfrac{x}{y} \bigg ) q (y) + {\cal O} \bigg(\frac{1}{P_3^2} \bigg) \,,
\label{matching_def}
\end{equation}
where $\tilde{q} \, (q)$ denotes a quasi-PDF (light-cone PDF) of a parton inside a hadron, and $C$ is the matching coefficient.
After performing a perturbative expansion of the LHS and the RHS of Eq. (\ref{matching_def}) in powers of $\alpha_s$, one can show that the first-order correction to the matching coefficient is 
\begin{equation}
C (x)= \delta(1-x) + \dfrac{\alpha_{s}C_{F}}{2\pi}  \bigg [ \widetilde{\Gamma} (x) - \Gamma (x) \bigg ] +  \dfrac{\alpha_{s}C_{F}}{2\pi} \delta (1-x)  \bigg [ \widetilde{\Pi}  - \Pi \bigg ] \,.
\label{Gamma_Pi}
\end{equation}
In Eq.~(\ref{Gamma_Pi}), $\Gamma$ ($\widetilde{\Gamma}$) and $\Pi$ ($\widetilde{\Pi}$) represent the real corrections and the virtual corrections for the light-cone (quasi-) PDFs, respectively. 
To obtain the light-cone PDF from such a perturbative matching, one needs to invert Eq.~(\ref{matching_def}). 
This inversion is applied to one-loop order and, thus, will reverse the signs of the prefactors of $\alpha_s$ in Eq.~(\ref{Gamma_Pi}), and the integral will run from $-\infty$ to $+ \infty$.
The fact that the matching formula in Eq.~(\ref{matching_def}), which so far has been explicitly checked for twist-2 PDFs only, does hold as well for the (quark-target) calculation of the twist-3 $g_T$ can be considered a nontrivial outcome of this study.

We next highlight some important points involved in the construction of the matching coefficient for $g_T (x)$, by using results for a nonzero gluon mass as IR regulator.
As a first step, we rewrite the sum of real and virtual corrections in a full plus-prescription form. 
Recall that for an arbitrary function $f(x)$, a plus-prescription is defined as $[f(x)]_{+} = f(x) -\delta (1-x) \int dy f(y)$. 
This format naturally captures the cancellation of $x=1$ divergences between the real and virtual corrections. 
For the light-cone $g_T$, Eqs.~(\ref{e:gT_1a_mg_LC}),~(\ref{e:gT_1b_mg_LC}), and~(\ref{e:self_energy_mg_LC}) give
\begin{eqnarray}
\label{eq:LC_gT_mg_total}
    \dfrac{\alpha_{s}C_{F}}{2\pi}  \bigg [ \Gamma (x) +  \delta (1-x) \, \Pi  \bigg ] &=& g_T^{\rm{(1a)}}(x)+g_T^{\rm{(1b)+(1c)}}(x) + 
    \delta(1-x) \left( g^{\rm{(2a)}}_{T} +g_T^{\rm{(2b)+(2c)}}\right) \nonumber \\[0.2cm]
    & = & \dfrac{\alpha_{s} C_F}{2\pi} \left[- \delta(x) + \frac{-x^2+2x+1}{1-x}\ln \frac{\mu^2}{x m_g^2} + (1-x)  \right]_+ ,
\end{eqnarray}
in the $\msbar$ scheme if a finite gluon mass is used as the IR regulator. 
Note that, for ease of notation, in Eq.~(\ref{eq:LC_gT_mg_total}) we have dropped the index ``UV'' in the subtraction scale $\mu_{\UV}$.

Extra care is needed when applying the plus-prescription for quasi-PDFs, especially outside the physical region.
In particular, terms like $1/(1-y)$ that are present in diagram~(2d) require special attention. 
In the following, through the specific example of this diagram, we outline the steps needed to write one-loop corrections for the quasi-PDFs in a full plus-prescription format. 
Focusing on the $y>1$ region, which is sufficient to convey the main idea, we find
\begin{equation}
g^{\rm{(2d)}}_{T, \rm Q} = 
- \int_1^\infty dy \left[g^{\rm{(1d)}}_{T, \rm Q} (y) + \dfrac{\alpha_{s}C_{F}}{2\pi} \frac{1}{y}\right] + \dfrac{\alpha_{s}C_{F}}{2\pi} \bigg ( \frac{1}{2}\ln \frac{\mu^2}{4 p_3^2} \bigg ) \,.
\end{equation}
Thus, we can write $g^{\rm{(2d)}}_{T, \rm Q}$ in terms of the integral of $g^{\rm{(1d)}}_{T, \rm Q}$. The plus-prescription is then built as
\begin{equation}
  g^{\rm{(1d)}}_{T, \rm Q}(x)+ \delta(1-x) \, g^{\rm{(2d)}}_{T, \rm Q} = \left[g^{\rm{(1d)}}_{T, \rm Q}(x) + \dfrac{\alpha_{s}C_{F}}{2\pi} \frac{1}{x}\right]_+ - \dfrac{\alpha_{s}C_{F}}{2\pi} \frac{1}{x} + \dfrac{\alpha_{s}C_{F}}{2\pi} \delta (1-x) \bigg ( \frac{1}{2}\ln \frac{\mu^2}{4 p_3^2} \bigg ) \,.
\end{equation}
Repeating the above steps for the other diagrams, we arrive at
\begin{eqnarray}
\label{eq:quasi_gT_mg_total}
\dfrac{\alpha_{s}C_{F}}{2\pi}  \bigg [ \widetilde{\Gamma} (x) +  \delta (1-x) \, \widetilde{\Pi}  \bigg ] &=& g_{T,\rm{Q}}^{\rm{(1a)}}(x) + g_{T,\rm{Q}}^{\rm{(1b)+(1c)}}(x) + g_{T,\rm{Q}}^{\rm{(1d)}}(x) + 
\delta(1-x) \left( g^{\rm{(2a)}}_{T, \rm Q} + g_{T,\rm{Q}}^{\rm{(2b)+(2c)}} 
+ g_{T,\rm{Q}}^{\rm{(2d)}}\right)   \nonumber \\[0.2cm]
&  & =
\frac{\alpha_s C_F}{2\pi}
\begin{dcases}
\left[\frac{-x^2 + 2x +1}{1-x} \, \ln \frac{x}{x - 1} - 1 + \frac{1}{1-x} + \frac{3}{2x}\right]_+ - \frac{3}{2x}
&  x > 1  \\[0.2cm]
\left[\frac{-x^2 + 2x +1}{1-x} \, \ln \frac{4(1 - x) p^{2}_{3}}{m_g^2} + 1 - 2x - \frac{1}{1-x}\right]_+
&  0 < x < 1 \\[0.2cm]
\left[\frac{-x^2 + 2x +1}{1-x} \, \ln \frac{x - 1}{x} + 1 - \frac{1}{1-x} + \frac{3}{2(1-x)} \right]_+ - \frac{3}{2(1-x)}
&  x < 0 \nonumber \\
\end{dcases} \nonumber \\
& & + \frac{\alpha_s C_F}{2\pi} \delta(1-x) \left(\frac{1}{2}+\frac{3}{2}\ln\frac{\mu^2}{4 p_3^2} \right) \,,
\end{eqnarray}%
in the $\msbar$ scheme. 

Finally, combining expressions~(\ref{eq:LC_gT_mg_total}) and~(\ref{eq:quasi_gT_mg_total}) as per Eq.~(\ref{Gamma_Pi}), we obtain the matching coefficient
\begin{eqnarray}
C_{\scriptsize \msbar}\left(\xi,\frac{\mu^2}{p_3^2}\right)  &=&\delta(1-\xi)  \nonumber\\ 
&+& \frac{\alpha_s C_F}{2\pi}
\begin{dcases}
\left[\frac{-\xi^2 + 2\xi +1}{1-\xi} \, \ln \frac{\xi}{\xi - 1}  + \frac{\xi}{1-\xi} + \frac{3}{2\xi}\right]_+ - \frac{3}{2\xi}
& \quad \xi > 1  \\[0.2cm]
\delta(\xi) + \left[ \frac{-\xi^2 + 2\xi +1}{1-\xi} \, \ln \frac{4\xi(1 - \xi) p^{2}_{3}}{\mu^2} +
\frac{\xi^2-\xi-1}{1-\xi}  \right]_+
& \quad 0 < \xi < 1 \\[0.2cm]
\left[\frac{-\xi^2 + 2\xi +1}{1-\xi} \, \ln \frac{\xi - 1}{\xi} - \frac{\xi}{1-\xi} + \frac{3}{2(1-\xi)} \right]_+ - \frac{3}{2(1-\xi)}
& \quad \xi < 0 \nonumber \\
\end{dcases} \nonumber \\
& + &  \frac{\alpha_s C_F}{2\pi} \delta(1-\xi) \left(- \frac{1}{2}+\frac{3}{2}\ln\frac{\mu^2}{4 p_3^2} \right) \,.
\label{e:MSbar_gT_mg_matching}
\end{eqnarray}
Note that in Eq.~(\ref{e:MSbar_gT_mg_matching}) we have transformed variables $x\rightarrow\xi$ in order to keep $x$ as the variable representing the momentum fraction of the parent hadron carried by the quark, that is $p^3 = x P^3$. 
We re-emphasize that the IR singularities of the light-cone PDF (see Eq.~(\ref{eq:LC_gT_mg_total})) and the quasi-PDF (see Eq.~(\ref{eq:quasi_gT_mg_total})) are the same. As pointed out before, this is a requirement and the foundation for building up any matching equation. 
Furthermore, we find the same matching coefficient for all three IR regulators. 
This applies, in particular, also to the zero-mode contribution, regardless of how we regulate the IR region of the $k_{\perp}$ integral of the expression in Eq.~(\ref{e:Dirac_delta}).

The problem with the $\msbar$ renormalized matching coefficient is that the convolution integral relating the light-cone PDF to the quasi-PDF is UV divergent. 
These divergences originate from the integrals of the real corrections.
To illustrate this point, we use DR, restrict ourselves to the $\xi > 1$ region, and then generalize the underlined technique for the other regions. 
Furthermore, we employ a $\varepsilon$ regularization for the additional divergence at $\xi =1$. 
Applying these techniques, we arrive at
\begin{equation}
 \int_1^\infty d\xi \left(\frac{p^3}{\mu}\right)^{- \epsilon_{\UV}}(\xi)^{-\epsilon_{\UV}} \left[\frac{\xi^2-2\xi-1}{(1-\xi)^{1+\varepsilon}}\ln\frac{\xi-1}{\xi}+\frac{\xi}{(1-\xi)^{1+\varepsilon}}\right]
 =  -\frac{1}{4} - \frac{3}{2\epsilon_{\UV}} - \frac{3}{4}\ln\frac{\rm {\mu}^2}{p_3^2} + \, ...\,,
\end{equation}
where the terms in $\varepsilon$ have not been written because they are irrelevant for the discussion and ultimately drop out when combining the real and virtual corrections. Using DR, we can thus perform these integrals, which are otherwise divergent.
We observe that the $\xi$-dependence in the real diagrams is such that, when Eq.~(\ref{e:MSbar_gT_mg_matching}) is convoluted with the quasi-PDF, we get an unbalanced divergence, as the convolution integrals are normal integrals in the parton momentum fraction $\xi$, and not DR integrals.
Therefore, in order to work with a finite matching, we follow the procedure proposed in Ref.~\cite{Alexandrou:2019lfo} and use the so-called modified $\msbar$ ($\mmsbar$) scheme, which amounts to subtracting the divergent logs by renormalizing the whole $\xi$ dependence outside the physical region,
\begin{equation}
  \left[\frac{\xi^2-2\xi-1}{1-\xi}\ln\frac{\xi-1}{\xi}+\frac{\xi}{1-\xi}\right] \rightarrow
  \left[\frac{\xi^2-2\xi-1}{1-\xi}\ln\frac{\xi-1}{\xi}+\frac{\xi}{1-\xi} + \frac{3}{2\xi}\right]_R - \frac{3}{4}\ln\frac{\mu^2}{p_3^2}\delta (1-\xi) \,,
\end{equation}
where $[...]_R$ means that the whole $\xi$-dependence for $\xi>1$ was renormalized at the renormalization scale $\mu$.
A similar expression can be computed for the $\xi < 0$ region. 
However, the $\xi$-dependence inside the physical region is left untouched.
Renormalizing the  $\xi$-dependence outside the physical region results in finite integrals, but the norm is not preserved due to the remaining finite factors, $\pm 1/2 + 3/2 \ln (1/4)$ ($\pm$ depending upon the IR regulator).
In the $\mmsbar$ scheme, we use as the renormalization condition that the integral of the renormalized matching is equal to its tree-level value conveyed through subtracting also the finite parts.
In other words, we construct a matching coefficient such that the whole $\alpha_{s}$ correction integrates to zero so that the norm of the PDF is preserved.
Doing these subtractions, the matching for $g_T$ in the $\mmsbar$ scheme is 
\begin{eqnarray}
C_{\scriptsize \mmsbar}\left(\xi,\frac{\mu^2}{p_3^2}\right)  &=&\delta(1-\xi)  \nonumber\\ 
&+& \frac{\alpha_s C_F}{2\pi}
\begin{dcases}
\left[\frac{-\xi^2 + 2\xi +1}{1-\xi} \, \ln \frac{\xi}{\xi - 1}  + \frac{\xi}{1-\xi} + \frac{3}{2\xi}\right]_+ 
& \quad \xi > 1  \\[0.2cm]
\delta(\xi) + \left[ \frac{-\xi^2 + 2\xi +1}{1-\xi} \, \ln \frac{4\xi(1 - \xi) p^{2}_{3}}{\mu^2} +
\frac{\xi^2-\xi-1}{1-\xi} \right]_+
& \quad 0 < \xi < 1 \\[0.2cm]
\left[\frac{-\xi^2 + 2\xi +1}{1-\xi} \, \ln \frac{\xi - 1}{\xi} - \frac{\xi}{1-\xi} + \frac{3}{2(1-\xi)} \right]_+ 
& \quad \xi < 0  \,.
\end{dcases} 
\label{e:MMSbar_gT_matching}
\end{eqnarray}
We point out that the $\delta(x)$ zero-mode contribution in Eq.~(\ref{e:MMSbar_gT_matching}) is related to $g_{T,{\rm Q}}$ at $x = \infty$.
Presently, this point is not accessible in a model-independent manner through calculations in lattice QCD, but one might attempt to approximately determine the quasi-PDF for $x \to \infty$ in pQCD.
Note also that partonic calculations of the type presented here are more complicated when using nonzero parton masses (especially a nonzero quark mass).
This applies even more so when trying to extend the matching calculation to two loops.
In that case, DR for the IR region may well be by far the best choice from a pragmatic point of view.

The extra subtraction in the $\mmsbar$ scheme was made with 
\begin{equation}
\label{e:Renorm_MMS}
Z^{\scriptsize \mmsbar} (\xi) = 1 - \frac{\alpha_{s} C_F}{2\pi}\frac{3}{2}\left(-\frac{1}{\xi}\theta(\xi-1) - \frac{1}{1-\xi}\theta(-\xi)\right)
- {\alpha_s C_F\over 2\pi}\delta(1-\xi) \left( -\frac{1}{2}+{3\over2}\ln\left(\frac{1}{4}\right) \right) \,.
\end{equation}
This renormalization factor is structured so that the matching coefficients in the $\msbar$ scheme and the $\mmsbar$ scheme are related as
\begin{equation}
    C_{\scriptsize \mmsbar} = Z^{\scriptsize \mmsbar} C_{\scriptsize \msbar} \,.
\end{equation}
In position space, the renormalization factor in Eq.~(\ref{e:Renorm_MMS}) reads
\begin{eqnarray}
\label{e:RenormPosition_MMS}
Z^{\scriptsize \mmsbar} (z\mu) &=& 1 - \frac{\alpha_{s} C_F}{2\pi} e^{i z \mu}   \left(-\frac{1}{2}+\frac{3}{2}\ln\left(\frac{1}{4}\right)\right) \nonumber \\[0.15cm]
&+& \frac{3}{2}\frac{\alpha_{s}C_F}{2\pi}\left(i\pi \frac{|z \mu|}{2 z \mu} - {\rm Ci}(z \mu) + \ln(z \mu) - \ln(|z \mu|)- i {\rm Si}(z \mu)\right) \nonumber \\[0.15cm]
&-& \frac{3}{2} \frac{\alpha_{s}C_F}{2\pi} e^{i z \mu} \left(\frac{2 {\rm Ei}(-i z \mu)- \ln (-i z \mu) + \ln (i z \mu) + i \pi {\rm Sign}(z \mu)}{2}\right) \,,
\end{eqnarray}
where ${\rm Ci}$ is the cosine integral, ${\rm Si}$ the sine integral, ${\rm Ei}$ the exponential integral, and ${\rm Sign}$ the sign function.
In the limit $z \rightarrow 0$, one has
\begin{eqnarray}
Z^{\scriptsize \mmsbar} (z \rightarrow 0) &=& 1 - \frac{\alpha_{s} C_F}{2\pi} \bigg ( -\dfrac{1}{2} + \dfrac{3}{2} \ln \bigg ( \dfrac{z^{2}\mu^{2}e^{2\gamma_{E}}}{4} \bigg ) \bigg ) \,.
\end{eqnarray}
Therefore, the renormalization condition in Eq.~(\ref{e:RenormPosition_MMS}) implies a cancellation of $\ln(z^{2})$ singularity present in the $\msbar$ scheme.

\section{Summary and outlook}
\label{s:summary}

In this paper, we have derived the one-loop matching coefficient which relates the twist-3 light-cone PDF $g_T (x)$ to the corresponding quasi-PDF $g_{T,{\rm Q}}(x)$.
Generally, this type of matching can be considered as a factorization theorem connecting light-cone distributions to Euclidean correlators, which are calculable in lattice QCD.
Here, we have scrutinized this factorization for the first time at the twist-3  level.
Our results have been obtained using three different IR regulators: nonzero gluon mass, nonzero quark mass, and DR. 
The UV divergences have been dealt with DR throughout. 
Most importantly, we have found that the IR singularities of $g_T$ and $g_{T,\rm{Q}}$ exactly match. 
This is an encouraging result, which clearly supports the idea that the quasi-PDF method (and related approaches) is not limited to twist-2 parton correlators. 

The finite terms for individual diagrams are generally (quite) different.
Yet, the final result for the matching coefficient, after summing over all diagrams, does not depend on the IR regulator, which is another essential outcome of this study.
A symptom of the twist-3 nature of the light-cone $g_T(x)$ is the presence of a $\delta(x)$ (zero-mode) contribution in the matching coefficient.
We also found that a nonzero gluon mass, strictly speaking, is not sufficient to regulate the IR divergence related to this contribution.
According to what is known at present, such zero-mode contributions do not exist at the twist-2 level.

The matching coefficient has been provided in the $\msbar$ scheme and in the $\mmsbar$ scheme, which originally had been introduced to deal with extra complications one encounters in the context of quasi-PDFs~\cite{Alexandrou:2019lfo}.
More specifically, results in the former scheme have divergences showing up when the hard matching kernel is convoluted with the quasi-PDFs calculated in lattice QCD. 
The $\mmsbar$ scheme systematically removes all potential divergences and is designed such that it preserves the norm of the PDF.
Recently, the DR matching coefficient obtained here has been used in the first computation of $g_T(x)$ in lattice QCD~\cite{Bhattacharya:2020cen}.
Results for the twist-3 PDFs $e(x)$ and $h_L (x)$ will be presented elsewhere~\cite{Bhattacharya:2020jfj}. 
While more work is needed at the twist-3 level, such as a careful study of potential mixing of both light-cone PDFs and quasi-PDFs under UV renormalization, the results presented here support the quasi-PDF approach as a viable tool for studying the $x$-dependence of twist-3 parton correlators in lattice QCD.

\begin{acknowledgements}
The work of S.B.~and A.M.~has been supported by the National Science Foundation under grant number PHY-1812359.~A.M.~has also been supported by the U.S.~Department of Energy, Office of  Science, Office of Nuclear Physics, within the framework of the TMD Topical Collaboration. 
K.C. and A.S.\ are supported by the National Science Centre (Poland) grant SONATA BIS no.\ 2016/22/E/ST2/00013.
M.C. acknowledges financial support by the U.S. National Science Foundation under Grant No.\ PHY-1714407. 
F.S.~was funded by DFG project number 392578569.

\end{acknowledgements}

\noindent

\bibliography{gT_matching.bib}

\end{document}